\documentclass{ws-p8-50x6-00}

\begin{document}

\title{OBSERVATIONAL TESTS OF INTERMEDIATE MASS STAR YIELDS USING PLANETARY NEBULAE}

\author{K.B. Kwitter}

\address{Department of Astronomy, Williams College, Williamstown, MA 01267, USA\\
E-mail: kkwitter@williams.edu}

\author{R.B.C. Henry}

\address{Department of Physics \& Astronomy, University of Oklahoma, Norman, OK  73019,
USA\\E-mail: henry@mail.nhn.ou.edu}


\maketitle

\abstracts{
This paper summarizes a project designed to study abundances in a sample of planetary nebulae representing a broad range in progenitor mass and metallicity. We collect abundances of C, N, and O determined for the entire sample and compare them with theoretical predictions of planetary nebula abundances from a grid of intermediate-mass star models. We find very good agreement between observations and theory, lending strong support to our current understanding of nucleosynthesis in stars with progenitor masses below 8~M$_{\odot}$. This agreement between observation and theory also supports the validity of published stellar yields of C and N in the study of the abundance evolution of these two elements.}

\section{Introduction}

The cores and shell-fusing regions of intermediate-mass stars (1$<$M$<$8~M$_{\odot}$) become hot enough to synthesize C and N. Planetary nebulae contain the synthesis products of intermediate-mass stars. We determined C and N abundances for 20 PNe and compared derived abundances with stellar model predictions for PNe with progenitors over this same mass range. Since these same models predict chemical yields for C and N, our analysis constitutes an observational check on the validity of these theoretical yields.

\section{Results \& Conclusions}

Abundances are determined using emission-line data from ground-based (KPNO) and orbiting (IUE) telescopes. Calculations using appropriate atomic data for reaction rates provide abundances of ions associated with observed emission lines. The sum of observed ionic abundances is multiplied by an ionization correction factor to account for the contribution of unobserved ions to the total element abundance. A second correction is made using photoionization models to produce the final abundances.

Figs.~1A,B compare our abundance results (filled circles) for C and N versus O with model predictions of van~den~Hoek \& Groenewegen\cite{vdhg} (solid lines) and Marigo et al.\cite{m} (dashed lines). Lines connect models of indicated progenitor mass over a range of metallicity. Modest agreement is shown.

Our overall {\it conclusions} are: (1)~C and N abundances in PNe show much more scatter than they do in H~II regions, implying that PN progenitors produce and expel these two elements; (2)~the abundances of C and N in our sample of PNe are consistent with model predictions of intermediate-mass stars; and (3)~our abundance patterns are consistent with the production of N via hot-bottom burning in intermediate-mass stars with progenitors above 3.5-4~M$_{\odot}$.

Our complete study is available in Henry, Kwitter, \& Bates\cite{hkb} and earlier references cited therein.

\begin{figure}[t]
\epsfxsize=14pc 
\epsfbox{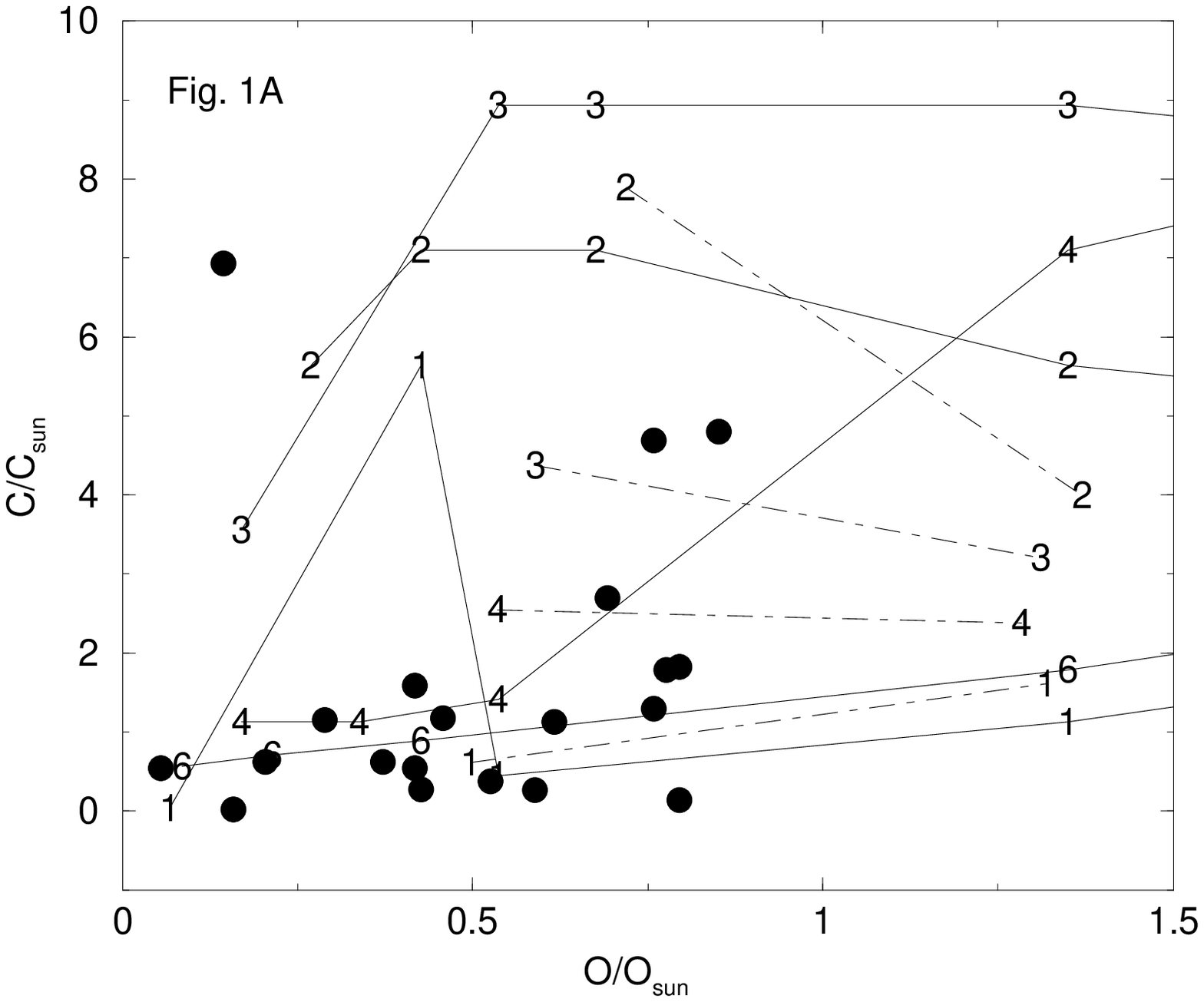} 
\epsfxsize=14pc 
\epsfbox{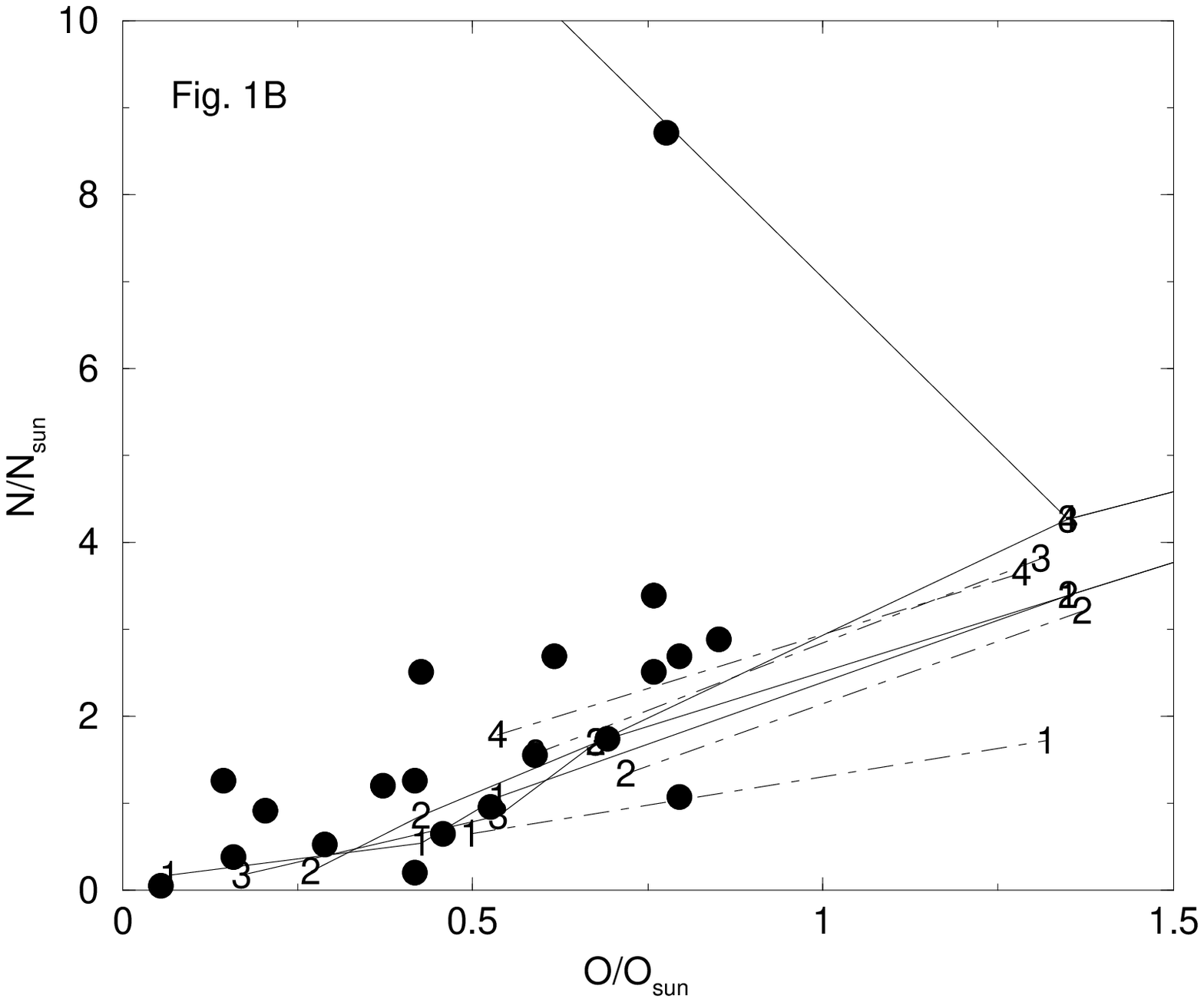} 
\end{figure}

\section*{Acknowledgments}

We are grateful for support from NSF grant AST-9819123 and NASA grant NAG 5-2389. In addition we appreciate the support from the staff of KPNO while carrying out the observing portion of this program.

\end{document}